\documentclass[]{article}       
\usepackage[]{geometry} 
\usepackage{graphicx}
\usepackage{cite}
\usepackage{picinpar}
\usepackage{amsmath}
\usepackage{url}
\usepackage{flushend}
\usepackage[latin1]{inputenc}
\usepackage{colortbl}
\usepackage{soul}
\usepackage{multirow}
\usepackage{pifont}
\usepackage{color}
\usepackage{alltt}
\usepackage[hidelinks]{hyperref}
\usepackage{enumerate}
\usepackage{siunitx}
\usepackage{breakurl}
\usepackage{epstopdf}
\usepackage{pbox}
\usepackage{algorithm}
\usepackage{algpseudocode}
\usepackage{amssymb}
\usepackage{subcaption}

\usepackage{palatino}

\providecommand{\keywords}[1]
{
  \small	
  \textbf{\textit{Keywords---}} #1
}

\begin{document}

\title{Knowledge Transferring via Model Aggregation for Online Social Care}

\author{Shaoxiong Ji$^{1}$,
        Guodong Long$^{2}$,
        Shirui Pan$^{3}$,
        Tianqing Zhu$^{2}$, \\
        Jing Jiang$^{2}$,
        Sen Wang$^{1}$,
        Xue Li$^{1}$ \\
        \small $^{1}${School of ITEE, Faculty of EAIT, The University of Queensland, Australia} \\ \small Email:~\{shaoxiong.ji;~sen.wang\}@uq.edu.au;~xueli@itee.uq.edu.au \\
        \small $^{2}${Centre for Artificial Intelligence, FEIT, University of Technology Sydney,  Australia} \\  \small Email:~\{guodong.long;~tianqing.zhu;~jing.jiang\}@uts.edu.au \\
        \small $^{3}${Faculty of Information Technology, Monash University, Australia} \\ \small Email:~shirui.pan@monash.edu
}

\date{}

\maketitle

\begin{abstract}
The Internet and the Web are being increasingly used in proactive social care to provide people, especially the vulnerable, with a better life and services, and their derived social services generate enormous data.
However, the strict protection of privacy makes user's data become an isolated island and limits the predictive performance of standalone clients. 
To enable effective proactive social care and knowledge sharing within intelligent agents, this paper develops a knowledge transferring framework via model aggregation. 
Under this framework, distributed clients perform on-device training, and a third-party server integrates multiple clients' models and redistributes to clients for knowledge transferring among users.  
To improve the generalizability of the knowledge sharing, we further propose a novel model aggregation algorithm, namely the average difference descent aggregation (AvgDiffAgg for short).
In particular, to evaluate the effectiveness of the learning algorithm, we use a case study on the early detection and prevention of suicidal ideation, and the experiment results on four datasets derived from social communities demonstrate the effectiveness of the proposed learning method.

\end{abstract}

\section{Introduction}
Proactive care is a kind of public service for healthcare and community assistance by connecting health organizations, social workers, and targeted patients. Traditional care service is based on face-to-face interaction in a certain place between general practitioners or social workers and people in need. Recently, with the help of online communication such as social networking services and private chatting, a new form of proactive online social service for mental health care has become available to online communities. Proactive social care provides people with early warning and support information to detect and relieve their mental disorders and social-related issues before their condition worsens.

Proactive social care for patients with mental disorders, especially depression and suicidality, is one of the most crucial services of social care in the modern society and has attracted worldwide attention. Mental health plays an important role in an individual's state of well-being. 
Mental health issues, such as depression, anxiety and post-traumatic stress disorder, have an adverse impact on people's daily life and health status. 
Untreated severe mental disorders could lead to suicidal ideation. 
According to WHO reports, around 300 million people suffer from depression\footnote{WHO fact sheets about mental disorders, available at \url{https://www.who.int/news-room/fact-sheets/detail/mental-disorders}}, and about 900,000 people commit suicide worldwide every year\footnote{Suicide rates, Global Heath Observatory (GHO) data, available at \url{http://www.who.int/gho/ mental_health/suicide_rates/en/}}. Moreover, these figures continue to increase in every country across the world. 

The traditional way to treat a mental health condition is psychological treatment, such as cognitive behavior therapy and interpersonal psychotherapy. This treatment relies heavily on health professionals such as general practitioners and psychiatrists. But current health services are not adequate to ensue effective treatment for such a huge number of potential sufferers. Furthermore, it is difficult to identify mental health issues at an early stage and take preventative action.

With the advances in the Web and mobile technology, mental health services are now using mobile devices to monitor a patient's health status and provide a platform for private communication within online communities to express mental stress. 
Conversation is one of the simplest and most effective ways to relieve an individual's mental disorders and even suicidal ideation.
Online text-based communication helps people to express their feelings and sufferings in their daily life and work, providing psycholinguistic clues for early detection. It also provides a possible channel for volunteers and social workers to respond to risky social posts and address a sufferer's  mental health issues through supportive comments.

There are many online platforms, forums and applications for chatting, peer support and early prevention, for instance, HealthfulChat, an online peer health support community, containing several chat rooms for mental health such as anxiety, bipolar and depression\footnote{Available at \url{http://www.healthfulchat.org/mental-health-chat-rooms.html}}; ReachOut Forums\footnote{\url{https://au.reachout.com}}, an anonymous space for 14 to 25 year old Australians to share stories and receive support online discussion; Turn2Me\footnote{\url{https://turn2me.org/}}, a lifeline web space for sharing and discussing personal issues; and 
Ibobbly, a mobile health intervention application \cite{tighe2017ibobbly}. Some services offered on these websites are delivered by mental health professionals. These services help people reach out and engage in conversations and consultation through the online communities. 

Text-based chatting services such as SMS services, mobile APPs, Web applications, and social networking services could also be helpful in promoting mental wellbeing when integrated with mental health care services. Several works have studied the text-based synchronous conversations for mental health intervention \cite{hoermann2017application}. Other works, especially those on early detection, focus on social networks for recognizing depression \cite{tsugawa2015recognizing}, detecting stress \cite{lin2017detecting}, and social network mental disorders \cite{shuai2018comprehensive}.
These early detection strategies are preliminary for proactive online social care services for social support.

However, user generated messages are isolated in some private communication platforms like SMS and private chatting room services, which leads to challenge for effective detection. Deep learning techniques benefit from large-scale training data. But in such isolated small datasets, the detection performance could be severely dampened. 
Human beings have the ability of learning knowledge by transferring within two different domains. 
Federated transfer learning (FTL) \cite{yang2019federated} provides an approach for knowledge transferring among federated agents in a decentralized manner. 
Following the principle of FTL, we developed a knowledge transferring framework for online social care by aggregating learned knowledge on distributed clients. Specifically, clients send their local models to a global server and the global server ensembles models by aggregation algorithms. The aggregated model on a global server acts as the shared knowledge of all the clients, and it can be redistributed to clients for knowledge transferring. 

This paper proposes a cooperative framework with knowledge transferring and a novel model aggregation method for online social care, together with several components for proactive care services. 
This solution comprises four key features, i.e., language representation, on-device training, mental health detection, and effectiveness stratification of supportive responses, to empower intelligent proactive social care for mental health. 
We develop deep neural networks to learn the representation of text for language understanding, and study two tasks to enable proactive online social care under the framework of data protection.

This paper contributes to the literature in the following three ways:
\begin{itemize}
\item We propose a cooperative knowledge transferring framework with model aggregation for proactive social care by introducing a third party model server for knowledge ensemble. 
\item To improve knowledge ensemble for effective knowledge sharing and transferring, we proposed a two-step optimization and average difference descent for model aggregation.
\item  To evaluate our proposed algorithm, a case study on suicidal ideation detection and effectiveness stratification as services of online social care is conducted, resulting in better performance than the baselines. 
\end{itemize}

The structure of this paper is as follows. Related work are reviewed in Section \ref{sec:related}. Our proposed framework is introduced in Section \ref{sec:method} together with an improved optimization algorithm. In Section \ref{sec:exp}, an experimental evaluation is conducted under the settings of proposed framework for online social care. A conclusion is drawn in Section \ref{sec:conclusion} together with a brief discussion.

\section{Related~Work}\label{sec:related}
This paper is related to mental health care, such as the detection of depression or suicidality and conversation treatment, and federated learning and its variant federated transfer learning.

\subsection{Mental Health Care}
A large body of research focuses on mental health to provide proactive care for those who need it, especially the detection of mental health issues such as stressor events \cite{li2017analyzing}, depression \cite{de2013predicting}, and suicidality \cite{o2015detecting}. Shuai et al. used a machine learning based model to perform multi-source learning for mental disorder detection in social media \cite{shuai2018comprehensive}.
Tsugawa et al. extracted features from a user's Twitter activities and detected that the user was suffering from depression \cite{tsugawa2015recognizing}.  Nguyen et al. performed affective and content analysis through a comparison between depression communities as the clinical group and normal communities as the control group \cite{nguyen2014affective}. Lin et al. proposed a hybrid method of factor graph and convolutional neural network to detect psychological stress through tweet content and user interaction \cite{lin2017detecting}.

Severe mental disorders could turn to suicidality. Suicidal risk has been studied from the perspective of interaction between clinicians and patients \cite{venek2017adolescent}, and knowledge discovery and detection using online social content \cite{ji2018supervised}. 
De Choudhury et al. investigated the transition of mental health to suicidality in online social communities \cite{de2016discovering}. Ren et al. proposed a complex emotion model for suicidal intention detection in blogs \cite{ren2016examining}. Ji et al. proposed an improved model aggregation method to detect suicidal ideation in a distributed manner \cite{ji2019detecting}.

\subsection{Federated Transfer Learning}

Federated learning \cite{mcmahan2017communication} is an on-device solution to decouple the training procedures from data collection. 
The method uses an iterative averaging model that can perform distributed training and learn efficiently from decentralized data to achieve the goal of preserving privacy. 
To improve communication efficiency,  Kone\v{c}n\`y et al. proposed structured updates and sketched updates to reduce uplink communication costs \cite{konevcny2016federated}. 
Geyer et al. proposed differential privacy preserving techniques on the client side to balance performance and privacy \cite{geyer2017differentially}.

Another scenario is federated transfer learning \cite{yang2019federated} was proposed to transfer knowledge in a federation with the combination of transfer learning \cite{pan2010survey} where decentralized agents have different samples and features. The learning principle is quite similar to fast adaptive meta-learning \cite{finn2017model, nichol2018firstorder} and zero-data learning \cite{larochelle2008zero}, that is, it learns a well-generalized global model in the data-free model server by aggregating the information learned from distributed clients.


\section{Method}\label{sec:method}
In this section, we develop a novel model aggregation over a federation of decentralized clients that enables knowledge transferring for proactive social care. It is designed with a third-party model server for knowledge ensemble and a two-step optimization strategy to decouple model training and data collecting. 
In particular, the datasets are located on decentralized clients, e.g. a physical electronic devices or an isolated software container, and the locally trained models serve as client's learned knowledge which are transported to global server for aggregation. Then, the global server aggregates the models into a global one as shared knowledge for clients. 

\subsection{Knowledge~Ensemble~and~Transferring}
\begin{figure}[htbp]
\begin{center}
\includegraphics[width=0.4\textwidth]{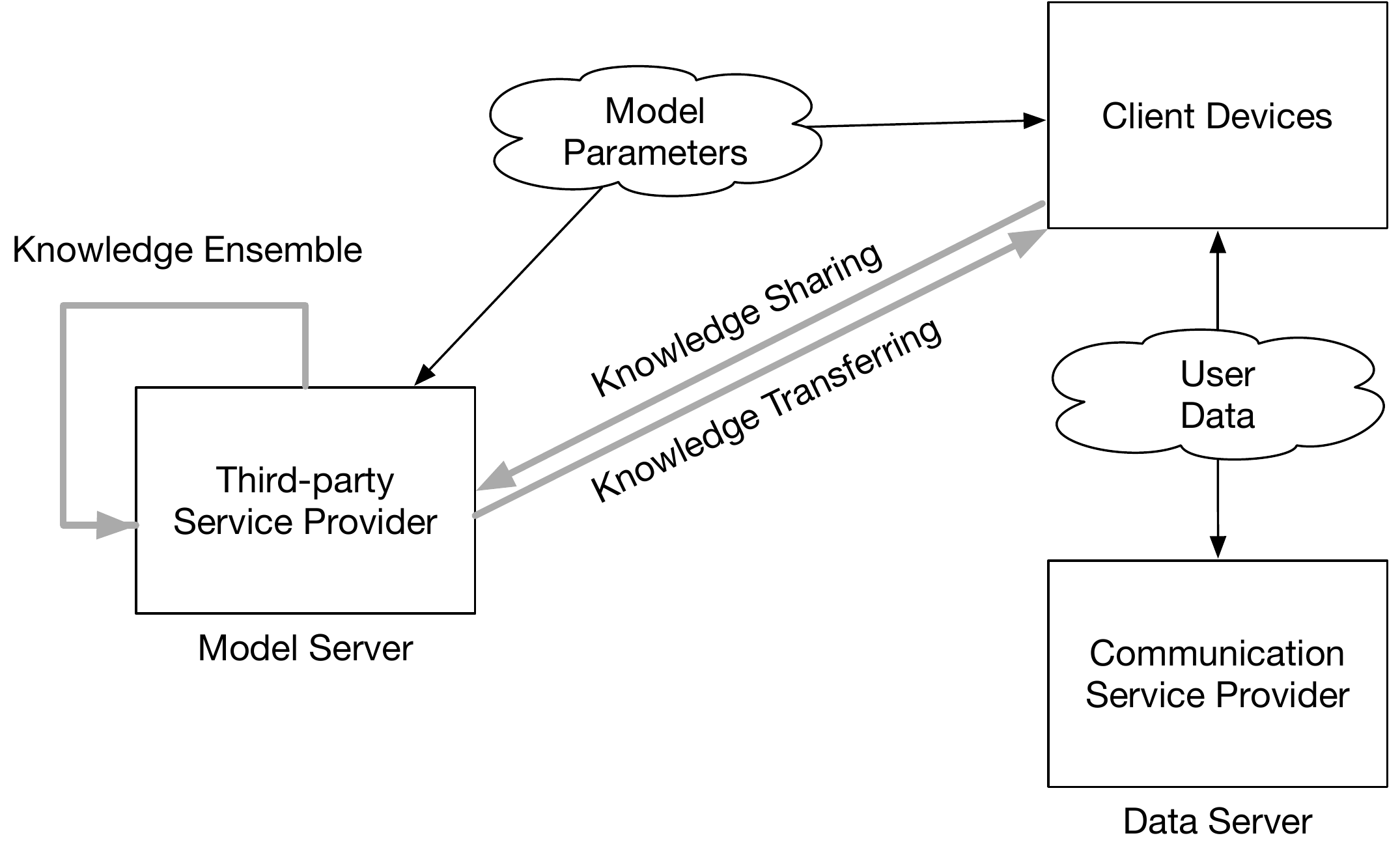}
\caption{The illustration of knowledge ensemble framework with sharing and transferring}
\label{fig:framework}
\end{center}
\end{figure}

Our method is under a decentralized learning framework by aggregation-based knowledge sharing and transferring for on-device training on local client devices. It is powered by a communication server with a chatting service for user data transmission and a third-party service provider, i.e., a mental health care service provider in this paper, for the communication of model parameters. The framework is illustrated in Figure \ref{fig:framework}, where the communication service is separated from third-party proactive social care, providing third-party applications with an approach to making inferences without accessing the raw data. Knowledge transferring is enabled via model aggregation and redistribution through client-server communication. 

\begin{figure*}[!ht]
\begin{center}
\includegraphics[width=0.9\textwidth]{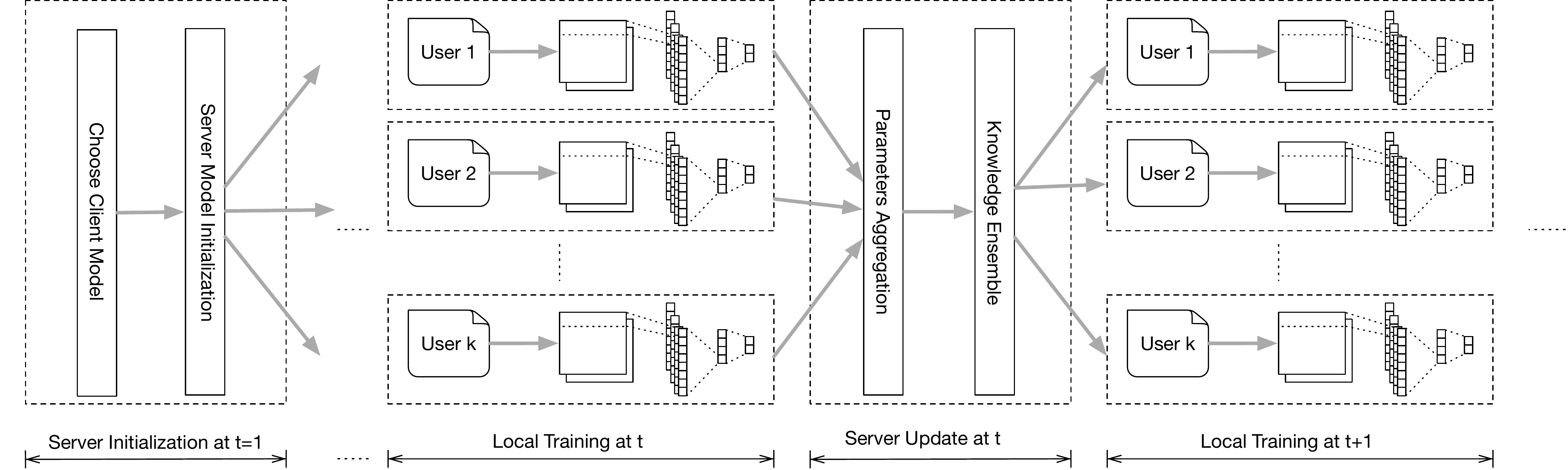}
\caption{The workflow of knowledge transferring framework with model aggregation when taking CNN as the learning model}
\label{fig:workflow}
\end{center}
\end{figure*}

The workflow of the knowledge transferring framework is illustrated in Figure \ref{fig:workflow}. First, the model server chooses a learning model as the client model for each client to perform specific tasks on devices. In this paper, we take two proactive social care tasks into consideration, i.e., text-based suicidality detection and social comment categorization. The first task aims to provide an early detection and warning system. The second task makes it easier for target users to access more effective responses. These two tasks are typically regarded as binary classification and multi-class classification problems, respectively. Deep neural networks, such as convolutional neural networks (CNNs) \cite{kim2014convolutional} for text and long short-term memory networks (LSTM) \cite{hochreiter1997long} are chosen as the classification model for clients to learn the language features in user generated content. In this illustration, CNN is used as an example. After parameter initialization, the model parameters as a form of shared knowledge are sent to the selected online clients to perform local training using the data of each user. Then, the locally trained models are sent back to the model server for model aggregation and knowledge updating. The intuitive way to undertake model aggregation is model averaging which takes local models equally. We propose a novel approach to aggregating a client model to optimize the global model for better knowledge transferring, called \textbf{Av}era\textbf{g}e \textbf{Diff}erence \textbf{Agg}regation or  {AvgDiffAgg} for short. The objective function and our proposed two-step optimization is introduced in detail in Section \ref{sec:obj} and \ref{sec:opt}. A round of training consists of local training on devices, parameter sending, and model aggregation on the model server. The learning framework works through client and server communication by an iterative update.

\subsection{Objective~Function~}\label{sec:obj}
In the private chatting scenario, deep learning methods are trained on user own data on their devices. However, isolated data is inadequate for training a deep learning model. To solve this problem, knowledge ensemble and transferring in our proposed framework aim to provide a good initialization to every user so that they can fine-tune a personalized deep learning model accordingly.

Training a deep learning model is a non-convex optimization task with many local optimal solutions or optimal points in the solution manifold. To address the local optimal problem in deep learning model training, there is an empirical assumption that if the initialization point of the model parameters is close to the global optimal point, arriving at the global optimal point or gaining a ``better" local optimal point is more likely if the model is fine-tuned. Here,  ``better" is compared to the average results with randomly selected initialization points. Therefore, the optimal initialization point $\theta$ should be 
\begin{equation}
\arg \min_{\theta} \mathbb{L} = \arg \min_{\theta} \sum_{k=1}^n \frac{1}{n} {L(\theta, \theta^k)}
\end{equation}
where $\theta^k$ is the optimal parameter solution for the $k$-th user, and $L$ is the loss function for measuring the distance between initialization point $\theta$ and each user's optimal point $\theta_k$. 

\begin{figure}[htbp] 
   \centering
   \includegraphics[width=0.5\textwidth]{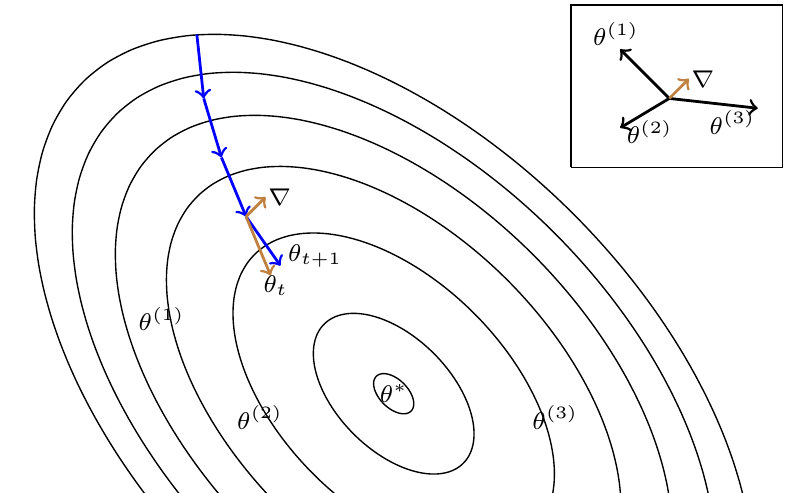} 
   \caption{Finding the nearest global weight to all the optimal local weights. The blue arrows show the model updating towards the optimum parameter $\theta^*$. The brown arrow $\nabla$ acts as the ``gradient''.}
   \label{fig:illustration}
\end{figure}

The procedure of finding the optimal global parameters is illustrated in Fig. \ref{fig:illustration}. The central body of this illustration shows how the server weights are updated to  optimal weights. The subfigure in the upper left corner shows how the local weights are composed as the gradient. The brown arrow in the form of the average difference between the model weights acts as the gradient. The optimization objective on the server side minimizes the average or expectation of the Euclidean distance between the server weight and the user weights. To facilitate this calculation, the loss function can be re-written as
\begin{equation}
    \label{eq:obj-func}
    \mathbb{L} = \sum_{k=1}^n \frac{1}{2n}L(\theta, \theta^k)^2
\end{equation}
where $L(\cdot,\cdot)$ is specified to Euclidean distance between two sets of weights, and $m$ is the number of users or local devices. 

\subsection{Two-step~Optimization}\label{sec:opt}
In Equation \ref{eq:obj-func}, the global model parameter $\theta$ and the $k$-th client model parameter $\theta_k$ are two correlated parameters that both  need to be optimized. To solve this optimization problem, we propose a two-step optimization algorithm that uses gradient descent to simultaneously approach optimal $\theta$ and $\theta_k$. Specifically, optimization is an iterative procedure and each iteration $t$ includes two steps that aim to separately update $\theta$ and $\theta_k$.

In the first step of each iteration, we clip the value of each user's parameters $\theta_k$, and update the global initialization point $\theta$ with the gradient derived from Equation \ref{eq:obj-func} as:
\begin{equation}
\frac{\partial\mathbb{L}}{\partial \theta} = \frac{1}{n} \sum_{k=1}^n (\theta_t - \theta_t^k),
\end{equation}
where $(\theta_t - \theta_t^k)$ is the difference between the  initialization point and the optimal point for the $k$-th user. In each iteration, we update the global initialization point with the average difference for all users, corresponding to Algorithm \ref{alg:diff} in this paper. 
\begin{equation}\label{eq:update}
    \theta_{t+1} \gets \theta_t - \epsilon \frac{1}{n}\sum_{k=1}^n(\theta_t-\theta_{t+1}^k)
\end{equation}
In practice, we can randomly sample part of the users in each iteration to estimate the ``average difference" so that we can reduce computation and avoid overfitting. As the estimation of $\theta$ only requires part of the users' parameters, the proposed optimization framework is robust for the scenario in which some users are disconnected during the training procedure.

In the second step of any iteration, the global initialization parameters $\theta$ are fixed as transferred knowledge to selected clients, and then we can fine-tune each user's $\theta_k$ with gradient descent by using the user's own data $D_k$:
\begin{equation}
\theta^k = O(\theta, D_k, F_k)
\end{equation}
where $O$ is an operator that iteratively updates the $\theta$ for a certain number of epochs, and $F_k$ is an arbitrary deep learning function applied on the local model of the $k$-th client. 

Once the local model's optimal parameters $\theta_t^k$ have been learned from the local model, they are sent to the central server to estimate the average difference which contributes to updating the global parameters $\theta_{t+1}$ for next-round knowledge tranferring.

\begin{algorithm}
\caption{Average Difference Descent for the Optimization of Knowledge transferring}
\label{alg:diff}
\begin{algorithmic}[1]
\State K is the total number of users; C is the fraction of users; $\epsilon$ is the stepsize of server execution.
\State \textbf{Input}: server parameters $\theta_{t}$ at~$t$, client parameters $\theta^1_{t+1}, ..., \theta^m_{t+1}$ at $t+1$.
\State \textbf{Output}: aggregated server parameters $\theta_{t+1}$.
\Procedure{Server~Optimization}{}
\State initialize~$\theta_0$

\For{each~round~$t$=1,~2,~...~}
\State $m\gets max(C \cdot K, 1)$
\State $S_t \gets (random~set~of~m~users)$
	\For{each~user~k~$\in~S_t$~on~local~device}
		\State $\theta_{t+1}^k \gets \text{LocalTraining}(k,\theta_t)$
	\EndFor
	\State $\theta_{t+1} \gets \theta_t - \epsilon \frac{1}{m}\sum_{k=1}^m(\theta_t-\theta_{t+1}^k)$ 
\EndFor
\EndProcedure
\end{algorithmic}
\end{algorithm}

\section{Experimental~Evaluation}\label{sec:exp}
In this section, we introduce the architecture of proactive social care in online communities. Two tasks for proactive service, i.e., suicidal ideation detection and social response categorization are studied. Datasets and baselines are introduced as well as a series of comparative experiments. 

\subsection{Online~Social~Care~}

\begin{figure}[htbp]
\begin{center}
\includegraphics[width=0.4\textwidth]{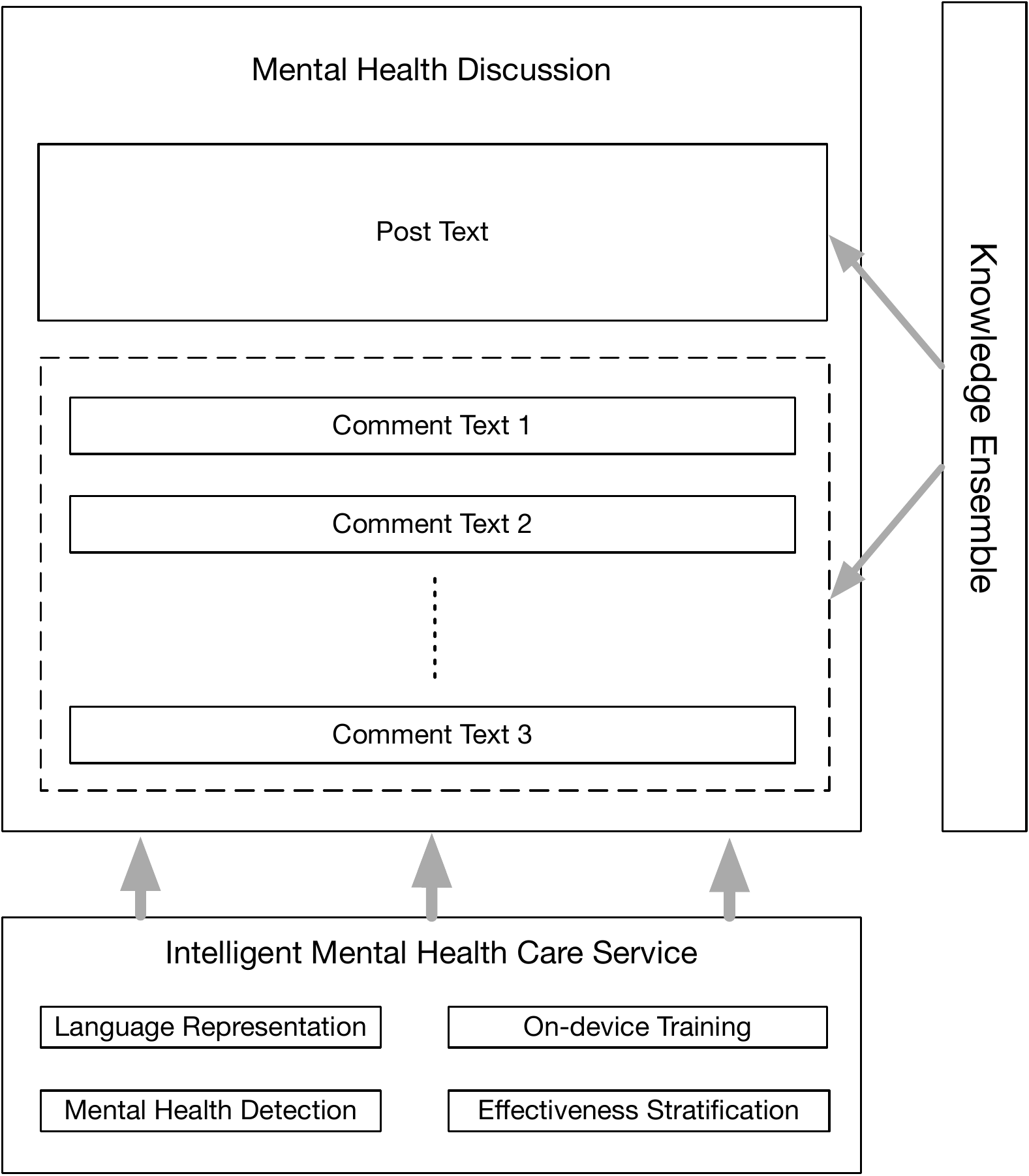}
\caption{The architecture of proactive social care for mental health}
\label{fig:architecture}
\end{center}
\end{figure}

Online social care provides many kinds of care services for targeted users. In this paper, we focus on mental health care in online communities for people who have a wide range of mental health issues. Under the learning framework, we produce two types of services, i.e., mental health detection and effectiveness stratification of social comments. We use suicidal ideation detection as the case study to demonstrate the application of proactive mental health detection. Suicide gestures and attempts are included in F60.3 -- Emotionally unstable personality disorder of ICD-10 code from WHO\footnote{\url{http://apps.who.int/classifications/apps/icd/icd10online2003/fr-icd.htm?gx60.htm+}}. Suicide is the most severe consequence of mental disorders. Post-schizophrenic depressive states may increase the risk of suicide\footnote{According to the ICD-10 code F20.4 -- Post-schizophrenic depression, available at \url{http://apps.who.int/classifications/apps/icd/icd10online2003/fr-icd.htm?gx60.htm+}.}. For effectiveness stratification, it provides an evaluation and ranking of people's comments and can be used for easy access to more persuasive comments. The architecture of proactive social care for mental health is illustrated in Figure \ref{fig:architecture}. We focus on the content from mental health discussion including the user's original post and the other user's comment on it. The proactive mental health care service is empowered by deep neural networks to learn language representation for early detection on posts and effectiveness stratification on comments. 

\subsection{Datasets}
We collected data from two social websites -- Reddit and Twitter. Table \ref{tab:datasets} lists the basic information of three datasets containing user posts derived from these platforms. 
For the task of effectiveness stratification, a dataset containing comment text is collected from Reddit. 

\begin{table}[ht]
    \centering
    \caption{Summary of datasets}
    \label{tab:datasets}
    \begin{tabular}{|c|c|c|}
    \hline
        Datasets & \# of users & \# of posts/tweets  \\ \hline \hline
        Reddit I & 99 & 39,600 \\
        Reddit II & 260 & 9,052 \\
        Twitter & 102 & 10,200 \\ \hline
    \end{tabular}
\end{table}

\noindent \textbf{Reddit~Dataset.~} We obtained two datasets from the website Reddit, which is ranked No. 6 on the list of top websites worldwide by Alexa\footnote{https://www.alexa.com/topsites} world wide as of June 2018. 
As a social website, Reddit aggregates a variety of topics for online discussions and each discussion community with an interest in a particular discussion is called a ``subreddit''. There are a wide range of topics for online discussion, including social events and personal experience.

The first aim of this work is to detect an individual's intent from social texts that involve suicidal ideation for early warning in proactive social care. 
A suicide-related subreddit called ``SuicideWatch''\footnote{\url{http://reddit.com/r/SuicideWatch}}, and two other subreddits not related to suicide, ``popular''\footnote{\url{http://reddit.com/r/popular}} and ``AskReddit''\footnote{\url{http://reddit.com/r/AskReddit}} are taken as the source of content with a total of 39,600 posts collected. Of these posts, there are 48.16\% of them containing suicidal ideation. 
We call this dataset as Reddit I.

Another dataset from Reddit, referred as Reddit II,  contains a total of 9,052 posts from a total of 260 selected users in the Reddit community.

\noindent \textbf{Twitter~Dataset.~} The third dataset was collected from the social website Twitter. A keyword filtering technique was applied to collect the original tweets.
 The filtering terms included words such as ``suicide'', ``die'', and ``death'', and suicide-related phrases, such as ``end my life'' and ``kill myself''. Then, we manually checked and labeled the posts. Tweets containing keywords but without suicidal ideation were put them in the control group. 
This Twitter dataset contains a total of 10,200 tweets, of which 5.8\% of tweets contained suicidal intention in the text.

\noindent \textbf{Reddit~Comments.~}
We collected comments from all the users in the Reddit II dataset for effectiveness stratification. Each comment had a score given by the users who had viewed the comment and clicked the ``like'' button in the online forum. We scaled the scores of the comments into five classes according to the score distribution. The number of comments on different posts varies.
Most posts contain less than 40 comments.

\noindent \textbf{Dataset~Partitioning.~} To mimic the real scenario of private chatting and decentralized training on the client, we partitioned the data using independently identical distribution (I.I.D.). First, a shuffle is applied to the entire dataset and it is partitioned into several users with a certain number of examples. There are 99 users and 102 users in the Reddit dataset and the Twitter dataset, respectively. Users of Reddit and Twitter had 400 posts and 100 tweets, respectively.

\subsection{Settings~and~Baselines}
To evaluate our model, two baselines with model aggregation, i.e.,  {FullbatchAgg} and {AverageAgg} (where Agg stands for \textbf{Agg}regation), are used for comparative experiments. 

These two baselines are described as follows:

\begin{enumerate}
    \item  {FullbatchAgg}: assembles an overall aggregation on the full batch of all users for only a single gradient descent step on each local device.
    \item  {AverageAgg}: samples a fraction of users for model aggregation applying weights over knowledge ensemble during model averaging.
\end{enumerate}

The  {FullbatchAgg} is a special case of  {AverageAgg} where the epoch of local training equals 1 and the fraction of users equals 1. 

For the learning models of clients, two popular deep neural models, i.e., CNN \cite{kim2014convolutional} and LSTM \cite{hochreiter1997long}, were used. First, we embedded the input sentence into a 100-dimension word vector to get the distributed representation of text. The word embedding was then placed into three convolutional layers. The learned features of the convolution layers are concatenated together to get the final representation of the text. Lastly, a fully connected layer was used as a classifier in the last layer to produce the prediction. For the LSTM model, we used the same settings for the word embedding and a 64-dimension LSTM hidden unit was used in the recurrent network.

\subsection{Suicidal~Ideation~Detection}
We firstly conduct experiments on suicidal ideation detection. To test the performance of our proposed learning framework and two-step knowledge transferring, we performed an empirical evaluation by comparing our method with those baselines.

%

\begin{figure}[ht] 
  \centering
  \begin{subfigure}[]{0.34\textwidth}
  	\centering
  	\includegraphics[width=\textwidth]{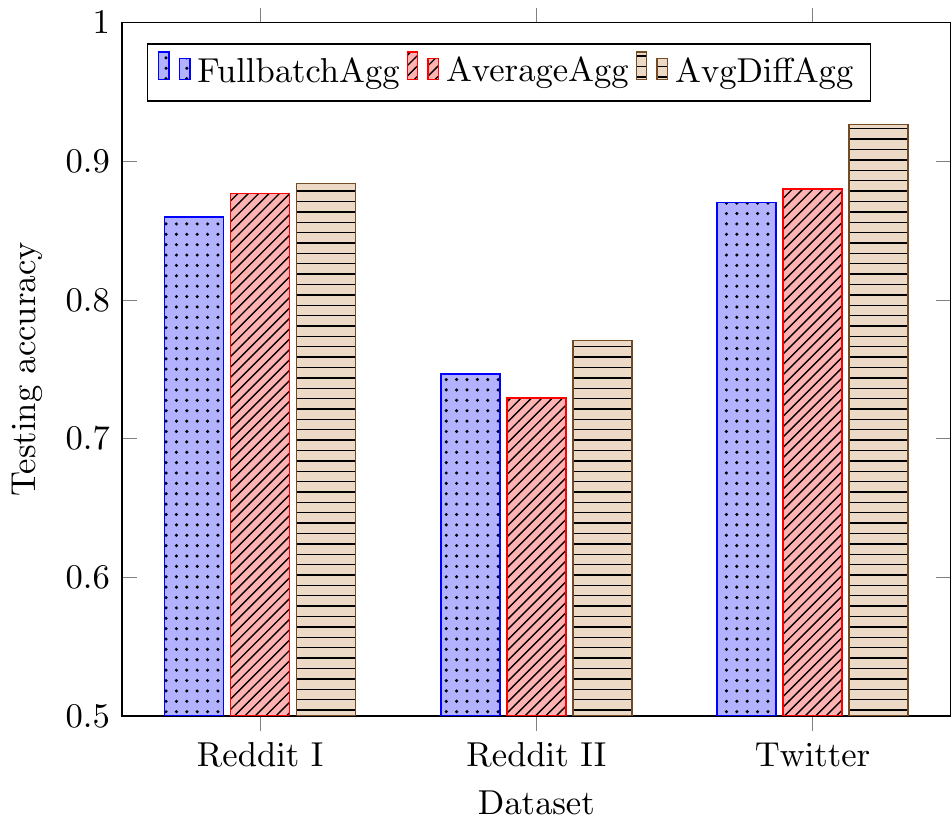} 
	\caption{Accuracy using CNN as the classifier}
  \end{subfigure}
  \begin{subfigure}[]{0.34\textwidth}
  	\centering
  	\includegraphics[width=\textwidth]{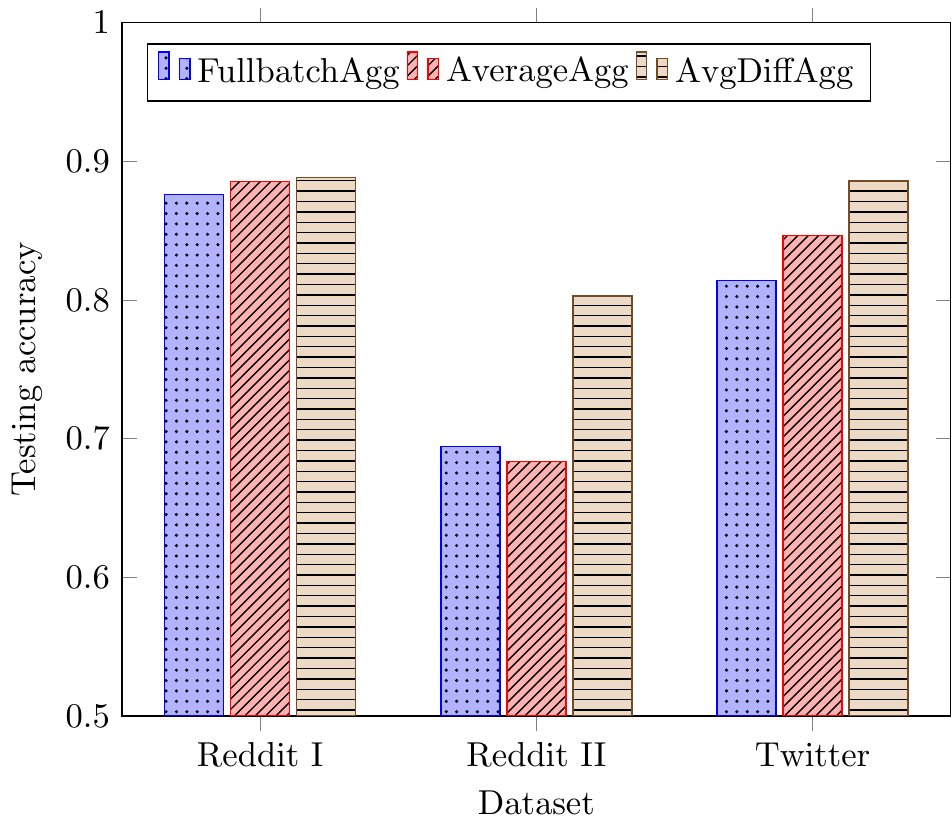} 
	\caption{Accuracy using LSTM as the classifier}
  \end{subfigure}
   \begin{subfigure}[]{0.34\textwidth}
  	\centering
  	\includegraphics[width=\textwidth]{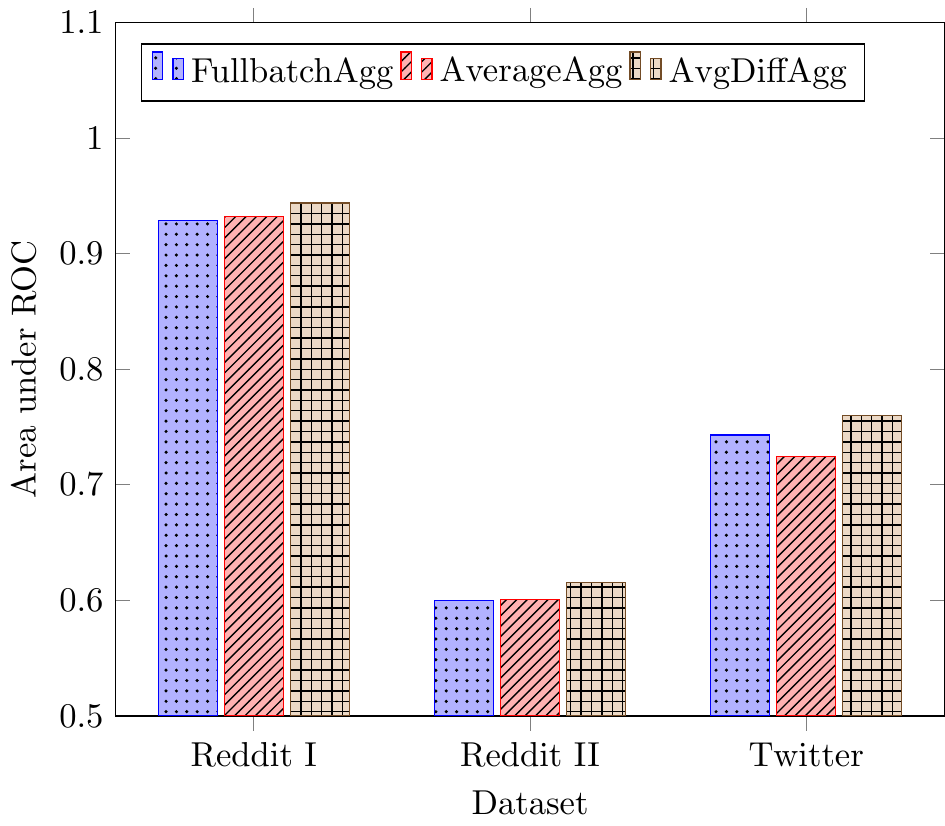} 
	\caption{AUROC using CNN as the classifier}
  \end{subfigure}
  \begin{subfigure}[]{0.34\textwidth}
  	\centering
  	\includegraphics[width=\textwidth]{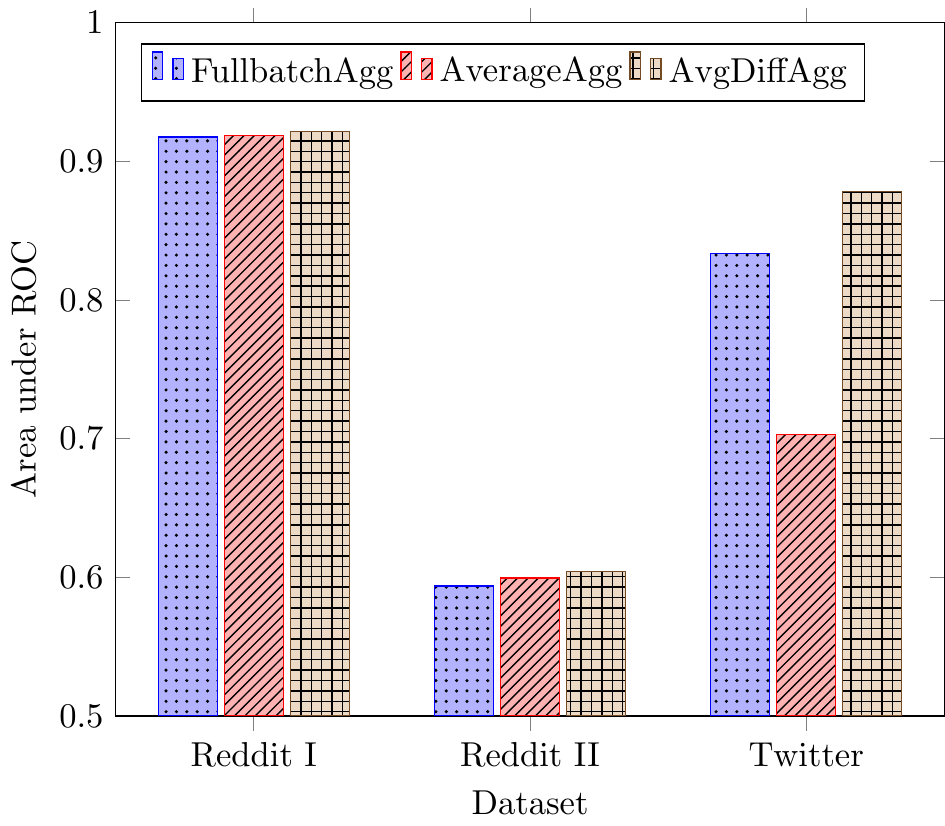} 
	\caption{AUROC using LSTM as the classifier}
  \end{subfigure}
\caption{Classification Accuracy and AUROC using CNN and LSTM as the classifier}
\label{fig:res-ldp}
\end{figure}

\vspace{1mm}
\noindent \textbf{Results.}
We compared all methods in terms of average testing accuracy and the average of area under the receiver operation curve (AUROC). The results are shown in Figure \ref{fig:res-ldp}. We used the same hyperparameter settings using the same number of training rounds of 10. The local batchsize was 10, and the local training epochs were 5. For  {AverageAgg} and our  {AvgDiffAgg}, the fraction of users was both set to 0.1. As we can see from these figures, our proposed method achieves the best scores when using both CNN and LSTM as the classifier.

\begin{figure}[ht] 
  \centering
  \begin{subfigure}[]{0.35\textwidth}
  	\centering
  	\includegraphics[width=\textwidth]{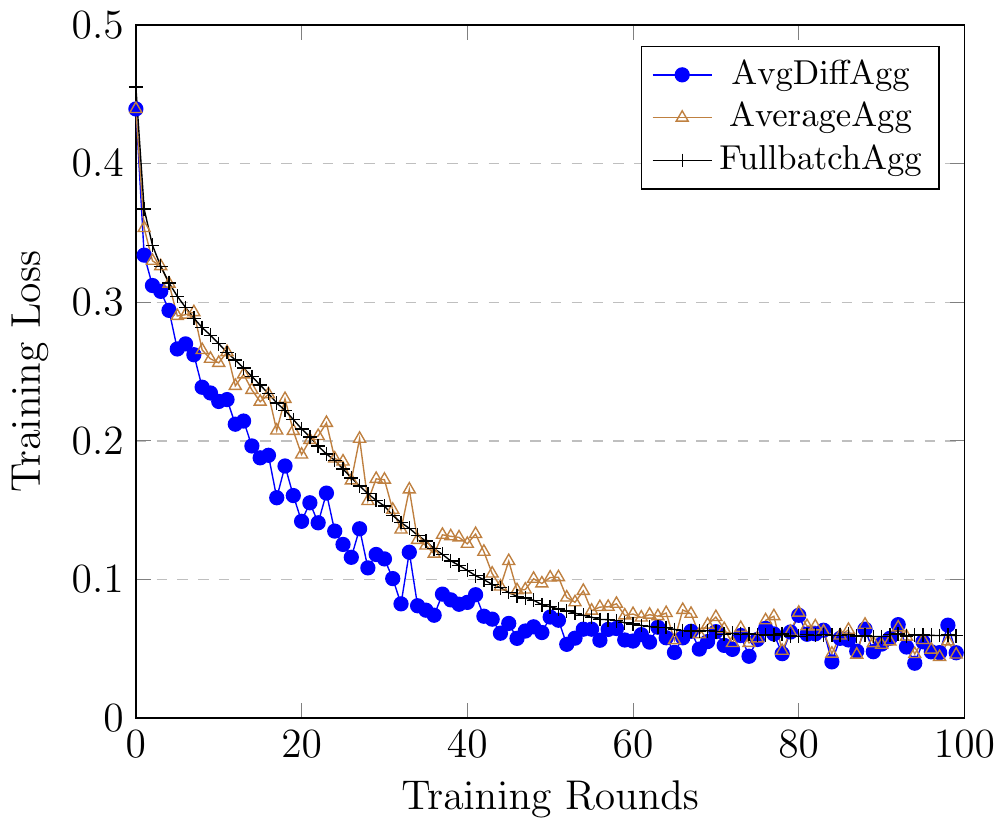} 
	\label{fig:loss}
	\caption{Training loss}
  \end{subfigure}
  \begin{subfigure}[]{0.35\textwidth}
  	\centering
  	\includegraphics[width=\textwidth]{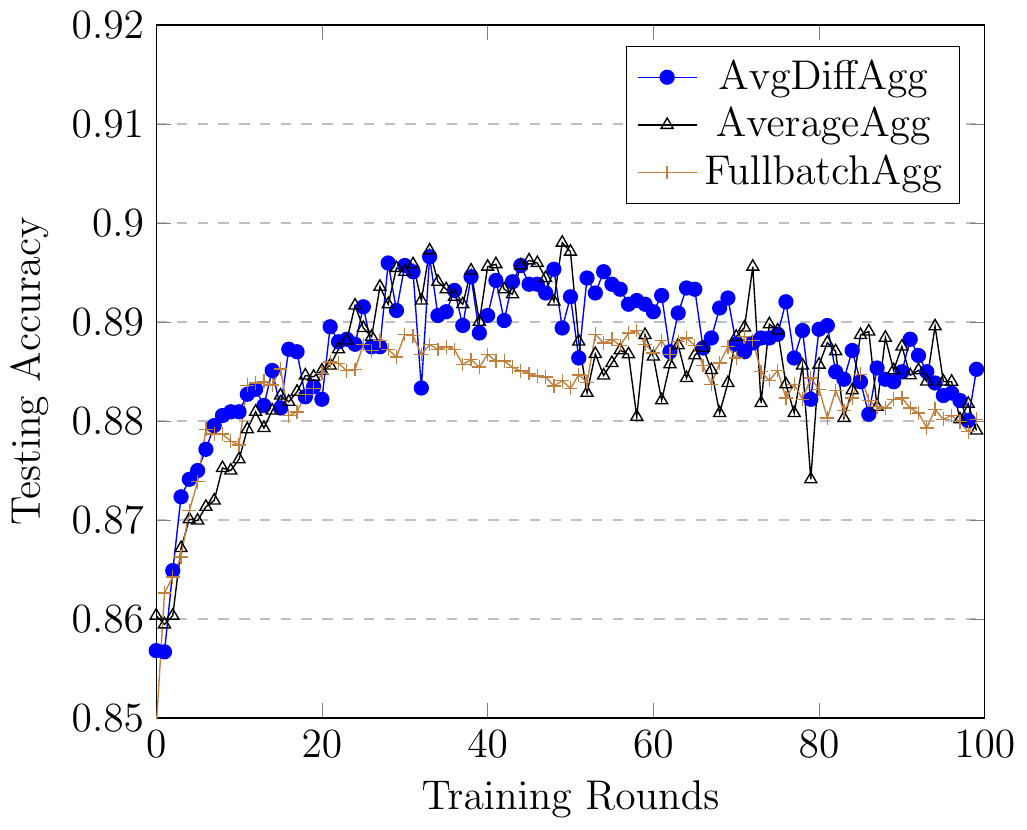} 
	\label{fig:acc}
	\caption{Testing accuracy}
  \end{subfigure}
\caption{Training loss and testing accuracy on Reddit
}
\label{fig:curve}
\end{figure}

\vspace{1mm}
\noindent \textbf{Learning curve.}
We drew the learning curve to visualize the performance of  {FullbatchAgg},  {AverageAgg}, and our  {AvgDiffAgg} as shown in Fig. \ref{fig:curve}. The training loss for our method decreased more rapidly than  {AverageAgg}. The test accuracy of our method was higher than  {AverageAgg} during the first 20 rounds of training and better still, between rounds 50 and 60. In the other rounds, the testing accuracy was similar. The training curve was smoother for the  {FullbatchAgg} because it uses the full batch of users during aggregation at each iteration, while the other two methods use a random selection of users for model aggregation. 

\subsection{Effectiveness~Stratification~of~Supporting~Words}
Conversation is one of the most effective ways to provide supportive words to the vulnerable people with mental health issues and even suicidal ideation.  
Gilat et al.\cite{gilat2012responses}, compared the responses to suicidal messages from trained volunteers and lay individuals, and found that trained volunteers employ more emotion-based strategies and more therapeutic-like cognitive-focused strategies than lay individuals who rely more on self-disclosure. 
The effectiveness stratification of supportive words evaluates social workers' responses in a given social care case, and it can help social workers to improve their conversational skill and compose better supportive words to persuade potential victims to relieve their mental health issue or give up a suicide attempt. 

In this section, we applied our  {AvgDiffAgg} and the same baseline methods to evaluate the potential effectiveness of social comments according to the score of each comment received from other users in a supervised way. 

We performed the experiments by training a CNN model and an LSTM model on the entire dataset with 10-fold cross validation. The average testing accuracy for the CNN and LSTM was 36.27\% and 35.33\%, respectively. These levels of accuracy are treated as the upper bound of the methods with data protection. The experiments using  {FullbatchAgg},  {AverageAgg} and  {AvgDiffAgg} were then performed 10 trials. The performance of different methods based on CNN and LSTM were then compared in terms of average testing accuracy. The experiment settings for these three methods were the same as the previous experiments, except for the fraction of users. For  {FullbatchAgg}, this was always 1, and for the other two methods, it was set to 0.1. The results are shown in Table \ref{tab:res-coms}. Methods without data protection had higher accuracy than methods with data protection. Of the methods with data protection, our proposed method using CNN as the classifier was slightly better than  {AverageAgg}. When using an LSTM as the classifier, the testing accuracy of our method was more than 2\% higher than  {AverageAgg}.

\begin{table}[ht]
\centering
\small
\caption{Comparison of accuracy on predicting comment scores as effectiveness stratification}
\label{tab:res-coms}
\begin{tabular}{|c|c|c| } \hline
\multirow{2}{*}{Methods} & \multicolumn{2}{c|}{Avg. Acc.} \\ \cline{2-3}
	&	CNN	&	LSTM	\\	\hline \hline
FullbatchAgg	& 30.49\%		& 31.67\%		\\	
AverageAgg	&	31.16\%	&	32.80\%	\\	
AvgDiffAgg	&	\textbf{31.35\%}	&	\textbf{35.08\%}	\\	\hline
\end{tabular}
\end{table}

\section{Conclusions}
\label{sec:conclusion}
Third-party intelligent web information systems could pave the way for effective social support and improve proactive online social care services from broad perspectives. To solve the scarcity of training data in decentralized settings, especially in the private chatting, this paper develops knowledge transferring method for proactive social care by using a decentralized learning framework with on-device model training, knowledge transferring, and novel model aggregation. In particular, the proposed model aggregation strategy for knowledge transferring updates the model parameters with the average difference descent that is inferred from a newly developed loss function customized for the proactive social service application scenario. The experiment evaluation on two tasks of suicidal ideation detection and effective stratification of social comments shows the effectiveness of the learning framework and the model aggregation algorithm.

Due to the highly sensitive nature of collecting real-world private data, this work mimics the real-world private chatting scenarios by using the public online data. The contents in the mimic dataset and the real-world private chatting dataset share similar characteristics and patterns that enable the proposed method to be a very promising solution to the development of new mental health care services in decentralized application. In future work, we will further research the real-world scenario and propose more efficient knowledge transferring methods while ensuring a comparable accuracy.

\bibliographystyle{plain} 
\bibliography{ref-social-care} 

\end{document}